\documentclass[iop, twocolumn, tighten]{emulateapj}
\usepackage{amsmath}

\newcommand{\Ndata}{N}

\newcommand{\Thetae}{\Theta}

\newcommand{\est}{{\mathrm{est}}}

\newcommand{\mestl}{{\boldsymbol m}_{\est,\lambda}}

\newcommand{\SVI}{S_\mathrm{VI}}

\newcommand{\eig}{\kappa}

\newcommand{\lambdar}{\lambda}

\newcommand{\er}{{\boldsymbol e}_R}

\newcommand{\elos}{{\boldsymbol e}_\mathrm{O}}
\newcommand{\es}{{\boldsymbol e}_\mathrm{S}}

\newcommand{\szeta}{\sin{\zeta}}
\newcommand{\czeta}{\cos{\zeta}}
\newcommand{\stheta}{\sin{\theta}}
\newcommand{\ctheta}{\cos{\theta}}

\newcommand{\trans}{\mathrm{T}}

\newcommand{\dv}{{\boldsymbol d}}

\newcommand{\error}{\varepsilon}
\newcommand{\errorv}{{\boldsymbol \error}}

\newcommand{\tdv}{\tilde{\boldsymbol d}}
\newcommand{\Dm}{G}
\newcommand{\tG}{\tilde{G}}

\newcommand{\mv}{{\boldsymbol m}}

\newcommand{\mpr}{\hat{{\boldsymbol m}}}

\newcommand{\ThetaS}{\Theta_{S}}
\newcommand{\omegaorb}{\omega_\mathrm{orb}}
\newcommand{\omegaspin}{\omega_\mathrm{spin}}

\slugcomment{}
\shortauthors{Kawahara and Fujii}
\shorttitle{Mapping clouds and terrain of exoplanets}
\begin{document}
\title{Mapping Clouds and Terrain of Earth-like Planets from Photometric Variability: Demonstration with Planets in Face-on Orbits}


\author{Hajime Kawahara\altaffilmark{1}
 and Yuka Fujii\altaffilmark{2}} 
\altaffiltext{1}{Department of Physics, Tokyo Metropolitan University,
  Hachioji, Tokyo 192-0397, Japan}
\altaffiltext{2}{Department of Physics, The University of Tokyo, 
Tokyo 113-0033, Japan}
\email{kawa\_h@tmu.ac.jp}
\begin{abstract}

We develop an inversion technique of annual scattered light curves to sketch a two-dimensional albedo map of exoplanets in face-on orbits. As a test-bed for future observations of extrasolar terrestrial planets, we apply this mapping technique to simulated light curves of a mock Earth-twin at a distance of 10 pc in a face-on circular orbit. A primary feature in recovered albedo maps traces the annual mean distribution of clouds. To extract information of other surface types, we attempt to reduce the cloud signal by taking difference of two bands. We find that the inversion of reflectivity difference between 0.8-0.9 and 0.4-0.5 $\mu \mathrm{m}$ bands roughly recover the continental distribution, except for high latitude regions persistently covered with clouds and snow. The inversion of the reflectivity difference across the red edge (0.8-0.9 and 0.6-0.7 $\mu \mathrm{m}$) emphasizes the vegetation features near the equator. The planetary obliquity and equinox can be estimated simultaneously with the mapping under the presence of clouds. We conclude that the photometric variability of the scattered light will be a powerful means for exploring the habitat of a second Earth.
\end{abstract}


\section{Introduction}
Recent discovery of rocky exoplanets provides hope for opening new doors for research of exoplanets which harbor life \citep[e.g.][]{2009A&A...506..287L,2011ApJ...729...27B}. Spectral biomarkers such as oxygen, ozone, water, methane \citep[e.g.][and references therein]{2002AsBio...2..153D,kaltenegger2010} and a characteristic feature of plants known as the red edge \citep[e.g.][]{2005AsBio...5..372S,2007AsBio...7..222K,2007AsBio...7..252K} will play a key role in searching for life on habitable planet candidates. It will also be important to understand the environment of the planetary surface, in other words, the habitat of the planet. Indeed, diverse surface environments on the Earth including continents, ocean, and meteorological condition serve as the backbone of biodiversity. One of the promising approaches to know the landscape of the terrestrial exoplanets is to identify surface components using the scattered light of the planets through the direct imaging observations. In this field, the Earth itself has been considered as a useful test-bed for future investigations for extrasolar terrestrial exoplanets. \citet{2001Natur.412..885F} demonstrated that the cloud distribution and inhomogeneous surface of the Earth generate photometric variability of scattered lights due to spin rotation. Thereafter, several authors attempted light curve inversion to characterize surface types \citep{2009ApJ...700..915C,2009ApJ...700.1428O,2010ApJ...715..866F,2011ApJ...731...76C}. 

One concern with the diurnal inversion is separation of clouds and other surface components since clouds can have an enormous contribution to total reflection light ($\sim$ 70 \% in energy for the Earth). The diurnal mapping with the single-band photometry is significantly polluted by the cloud reflection \citep{2009ApJ...700.1428O}. Diurnal inversion techniques with multiband photometry have been proposed to separate surface types, through the PCA analysis with 7 bands \citep{2009ApJ...700..915C,2011ApJ...731...76C} and the albedo modeling with 5 bands \citep{2011arXiv1102.3625F}.  
 
 The surface area facing on the observer and illuminated by the host star changes according to spin rotation and orbital revolution in different ways. We demonstrated that the two-dimensional mapping using the diurnal and annual multi-band photometric variations is possible for a cloud-free Earth under the assumption of the albedo model \citep[][hereafter KF10]{2010ApJ...720.1333K}. 
For the inversion with the diurnal and annual light curves, the cloud signal could be serious because seasonal variation of clouds is larger than diurnal variation \citep[e.g.][]{2008ApJ...676.1319P}. In this paper, we improve our previous method of the two-dimensional mapping to be applicable to a cloudy planet by using a single band or a few bands without any assumptions of albedo models. This problem is basically ill-posed and needs inversion technique used in the field of tomography. Hence we name this mapping method ``spin-orbit tomography''. We test this method by applying to light curves of a mock cloudy Earth-twin based on real satellite data. 

Another virtue of the two-dimensional mapping is the planetary obliquity measurement as discussed in KF10 for the cloud-free case. The obliquity is one of the few observables that have potential to constrain the formation scenario of the terrestrial planets \citep[e.g.][]{2007ApJ...671.2082K}. We also discuss the obliquity measurement for a cloudy planet. 
\section{Spin-Orbit Tomography}
\begin{figure}[htb]
  \includegraphics[width=\linewidth]{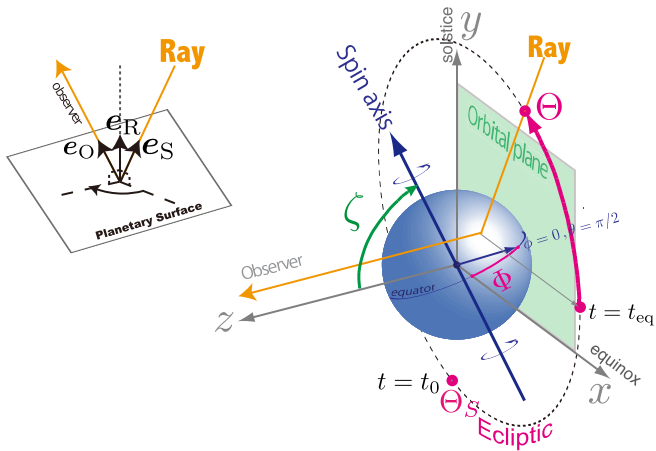}
\caption{Schematic configurations of the planetary system in a face-on orbit. The orbit lies in the $x$-$y$ plane, corresponding to the directions of the equinox ($x$-axis) and the solstice ($y$-axis) of the planet. The $z$-axis is defined by both the normal of the orbital plane and the direction of observers. The obliquity $\zeta$ is defined as the angle from the $z$-axis to the spin axis and the line of sight, respectively. The ecliptic longitude measured from the equinox is denoted by $\Theta$. We define $\ThetaS$ by the ecliptic longitude at the reference time $t = t_0$ so as to describe $\Theta = \omegaorb (t-t_0) + \ThetaS$. \label{fig:geo} }
\end{figure}

In general, scattered lights from the planetary surface are characterized by the bidirectional reflectance distribution function (BRDF), which is a function of three angles among the incident ray, the line of sight, and the normal direction of surface. Our inversion assumes the Lambert (isotropic) reflection to approximate real BRDFs of the surface. On this approximation, the intensity of the scattered light is expressed as
\begin{eqnarray}
\label{eq:in}
I_b =  \frac{R^2 F_{b}}{\pi} \int_{\SVI} W(\phi,\theta,\Theta,\Phi; \zeta, \ThetaS) m (\phi,\theta; b) \sin{\theta} d \theta d \phi , \nonumber \\
\end{eqnarray}
where $R$ is the planetary radius, $F_b$ is the incident flux of photometric band $b$, and $m(\phi,\theta; b)$ is the albedo of band $b$ at the planetary colatitude $\theta $ and longitude $\phi$. In this paper, we only consider a face-on circular orbit since the anisotropic scattering does not affect the phase curve. Figure \ref{fig:geo} displays the schematic configuration of the system. We denote the spin motion by the angle between  ($\phi=0$,$\theta=\pi/2$) and the direction of equinox, $\Phi$ and orbital revolution by the ecliptic longitude of the host star, $\Theta$ measured from the direction of equinox. Planetary obliquity $\zeta $ and the ecliptic longitude at the reference time $t_0$, $\ThetaS$ specify the spin axis. The $\ThetaS$ is related to Equinox Day as $t_\mathrm{eq} - t_0 = ( 2 \pi n - \ThetaS )/\omegaorb $ where $\omegaorb$ is the orbital angular velocity. The weight function is defined as $W(\phi,\theta;\Phi;\Thetae; \zeta, \ThetaS) \equiv (\es \cdot \er) (\elos \cdot \er)$ where $\es$, $\elos$, and  $\er$ are unit vectors from the planetary surface to a host star, from the surface to an observer, and from the planetary center to the surface, respectively. The integration is performed over the illuminated and visible area $\SVI$ defined as the region satisfying $\es \cdot \er > 0$ and $\elos \cdot \er >0$.  Taking the coordinate as shown in Figure \ref{fig:geo}, one obtains $\elos = (0,0,1)$, $\es = (\cos{\Thetae}, \sin{\Thetae}, 0) $ and $ \er = (\cos{(\phi+\Phi)} \stheta, \czeta \sin{(\phi+\Phi)} \stheta +  \szeta  \ctheta, -\szeta \sin{(\phi+\Phi)} \stheta + \czeta \ctheta )$. We set $R=1$ though an observable is $R^2 m(\phi,\theta; b)$. We use the light curve normalized by $F_b$, $ d (t; b) \equiv \pi I_b (t)/F_{b}$. In reality, $F_{b}$ can be estimated from the orbital distance of the planet, the observed host star flux, and the distance between the system and the Earth.

By discretizing the planetary surface $(\phi_j, \theta_j)$,  the light curve $d_i =  d (t_i; b)$ with observational noise $\error_i$ ($i=1,2,...,N$) can be modeled as
\begin{eqnarray}
\label{eq:dGm}
\dv &\approx& \Dm (\zeta,\ThetaS) \, \mv + \errorv, \\
  \Dm_{ij} (\zeta,\ThetaS) &\equiv& \left\{
\begin{array}{l}
\displaystyle{ W(\phi_j,\theta_j,\Theta(t_i),\Phi(t_i); \zeta, \ThetaS) \Delta \omega_s} \\
\;\;\;\;\;\;\mbox{ for $(\phi_j,\theta_j) \in \SVI $, } \nonumber \\
\displaystyle{0} \;\;\;\;\; \mbox{elsewhere}, \\
\end{array} \right. 
\end{eqnarray}
where $\Delta \omega_s$ and $m_j = m(\phi_j,\theta_j;b)$ are the solid angle and albedo of the $j-$th surface pixels ($j=1,2,...,M$). 
 In this paper, we use the HEALPix with $M=768$ for the surface pixelization \citep{2005ApJ...622..759G}.

We simultaneously solve $\mv$, $\zeta,$ and $\ThetaS$ from $\dv$ using the Tikhonov regularization \citep[e.g.][]{1989gdad.book.....M,tarantola,hansen}, which balances between observational noise and spatial resolution of the surface. The misfit function of the Tikhonov regularization is given by 
\begin{eqnarray}
Q_\lambda \equiv \sum_{i=1}^{\Ndata} \frac{|d_b(t_i) - \Dm_{ij} m_j |^2}{\sigma_i^2}  + \lambdar^2 |\mv - \mpr|^2,
\label{eq:qminimization}
\end{eqnarray}
where $\sigma_i$ is the standard deviation of the noise. We adopt the average of data as a prior of the model $\hat{m}_i = \langle d_b \rangle$. From the Bayesian viewpoint, the minimization of $Q_\lambda$ is regarded as maximization of a posteriori likelihood assuming a Gaussian prior for $\mv$ with the average $\mpr$ and the covariance matrix $\lambdar^{-2} I$, and an uniform prior for $\zeta$ and $\ThetaS$ \citep[e.g.][]{tarantola}. 
The solution which minimizes $Q_\lambda$ for given $\zeta$ and $\ThetaS$ is expressed as 
\begin{eqnarray}
\label{eq:tikres}
\mestl &=& V \Sigma_\lambdar U^\trans (\tdv - \tG \, \mpr) + \mpr \\
(\Sigma_\lambdar)_{ij} &\equiv& \frac{\eig_i}{\eig_i^2 + \lambdar^2} \delta_{ij}
\end{eqnarray}
where we define $\tilde{d}_i \equiv d_i/\sigma_i$ and $\tG_{ij} \equiv G_{ij}/\sigma_i$.  The orthogonal matrices $U$ and $V$ are given by the singular value decomposition of $\tG = U \Lambda V^\trans$ and $\eig_i$ is the $i$-th eigenvalue of $\tG$, that is, the $i$-th component of the diagonal matrix $\Lambda$ \citep[e.g.][]{hansen}. The $\delta_{ij}$ is the Kronecker delta. We search for the best-fit $\zeta$ and $\ThetaS$ by the Nelder-Mead method for different $\lambda$. Finally, the regularization parameter $\lambdar$ is determined by the L-curve criterion, which is the maximum curvature point of the model norm $|\mestl - \mpr|$ versus residuals $|\dv - G \mestl|$ plot \citep{hansen}.

\section{Results}
Here we demonstrate the spin-orbit tomography by applying it to simulated light curves of an Earth-twin. The light curves were computed with a line-by-line radiative transfer code RSTAR6B \citep{nakajima1988, nakajima1983} in combination with empirical data of monthly land reflectivity, monthly snow cover fraction, and daily cloud distribution in 2008 which are provided by Terra/MODIS \citep{MODIS}. 
The effect of multi-scattering, molecular lines, and the anisotropy of the scattering are included by RSTAR6B. We put cloud layers made of pure liquid water at an altitude of 4-6 km. We note that the different altitude assumptions (0-3 and 7-20 km) do not change the reflectivity of clouds in the three bands significantly ($<$5 \%). The cloud property is specified by two parameters, cover fraction and optical thickness. The reflection by ocean is modeled by the Fresnel reflection of a wavy ocean \citep{nakajima1983}. The detailed description is found in \citet{2011arXiv1102.3625F}.

\begin{figure*}[htb]
  \includegraphics[width=\linewidth]{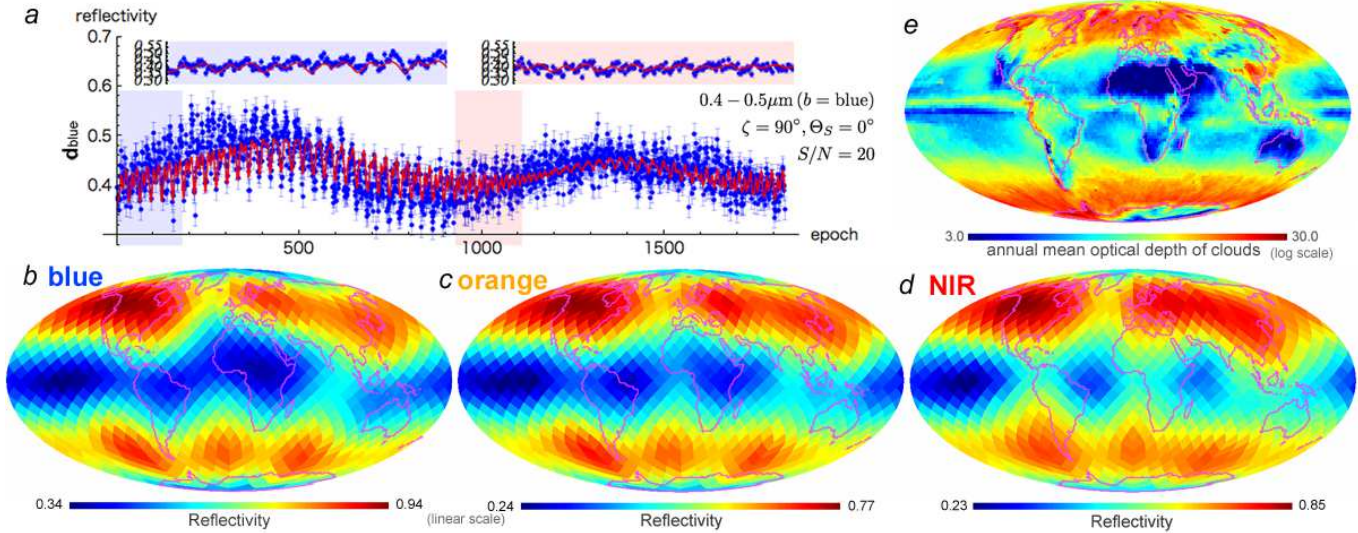}
\caption{The panel a displays the mock light curve of the blue band ($0.4-0.5 \mu \mathrm{m}$) of an Earth twin at a distance of 10 pc with a 4 m telescope and a hypothetical detector.  The data during 366 days are stacked for 6 days. Hence there are $30 \times 61 = 1830 $ data points.  Wider versions of two shaded regions are inserted in the mini panels. The predicted curve is drawn by solid lines. The recovered maps on the Mollweide projection are shown in the panel b (blue: $0.4-0.5 \mu \mathrm{m}$), c (orange:  $0.6-0.7 \mu \mathrm{m}$), and d (NIR: $0.8-0.9 \mu \mathrm{m}$). Magenta curves outline the boundary of continents and oceans. We fix $\zeta$ and $\ThetaS$ to the input values $90^\circ$ and $0^\circ$, respectively. The panel e displays annual mean of optical depth of clouds computed from the MODIS level 3 Daily Joint Aerosol/Water Vapor/Cloud Product in 2008 \citep{MODIS}. The broad bands of clouds at high latitudes correspond to the polar front and the narrow band of clouds at the equator is known as Intertropical Convergence Zone. \label{fig:sp}  }
\end{figure*}

\begin{figure}[htb]
  \includegraphics[width=\linewidth]{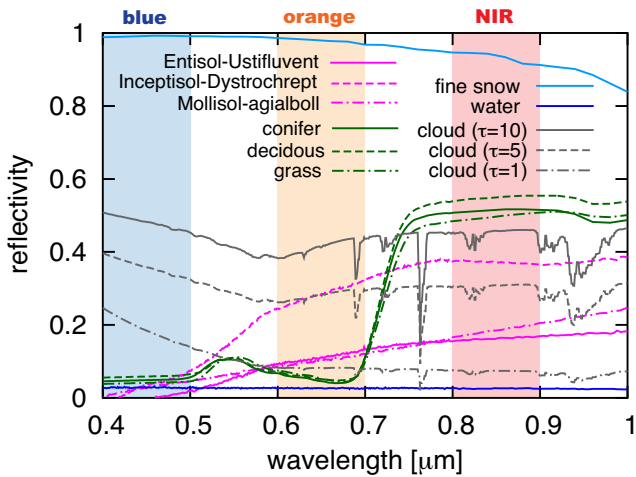}
\caption{Representable reflection spectra of clouds computed by {\it RSTAR6B} (gray), soil (magenta), vegetation (green), snow (cyan) and water (blue) taken from ASTER spectral library. The Entisol-Ustifluvent, Inceptisol-Dystrochrept, and Mollisol-agialboll correspond to ``Brown to dark brown silt loam '', ``Dark yellowish brown micaceous loam'', and ``Dark grayish brown silty loam'', respectively. Shaded regions indicate three bands we use.
 \label{fig:al}}
\end{figure}

In this paper we focus on photometric variability in three bands: $b=$ NIR (0.8-0.9 $\mu $m), orange (0.6-0.7$\mu $m), and blue (0.4-0.5$\mu $m) since these bands have large differences for soil and vegetations. These bands correspond to three of the bands of the EPOXI mission, which was used for validation of the simulation \citep{2011arXiv1102.3625F}. For demonstration purpose, we assume a face-on circular orbit and adopt $\zeta = 90^{\circ }$, which is the best geometry for a face-on orbit (KF10). Though this is not the case for general inclination, we postpone this problem until a forthcoming paper. 

We computed the synthetic scattered light of the Earth for 30 epochs per day (in 0.8 hour interval) over 366 days and stacked them for 6 days ($N = 61 \times 30$). Since the spin rotational period can be measured through periodogram analysis \citep{2008ApJ...676.1319P}, we assume that the $\omegaspin$ is known as well as $\omegaorb$. The Gaussian noise is imposed so as to mimic the observation with 
signal-to-noise ratio of $20$ for each bin during $0.8\times 6 =4.8$ hours, corresponding to a one year observation of an Earth twin at a distance of 10 pc with a 4 m telescope and a hypothetical detector (KF10). We estimate observational noises assuming a future mission with the occulter system, such as {\it the occulting ozone observatory} \citep[O3; e.g.][]{kasdin2010,2010SPIE.7731E..76S}, including the exzodiacal light (23 $\mathrm{mag \, arcsec^2}$), dark noise ($10^{-3} \mathrm{counts \, s^{-1}}$), read noise (2 $\mathrm{\sqrt{counts}/read}$), quantum efficiency (0.91), and end-to-end efficiency (0.5) with the parameters used in KF10 \citep[see also][]{2010PASP..122..401S}. While the starlight suppression ratio is difficult to discuss in general, a typical suppression ratio of an occulter is below $10^{-10}$ \citep[e.g.][]{2006Natur.442...51C}, even $10^{-12}$ for O3 \citep{2010SPIE.7731E..76S}. Other noises we assume are comparable to planet's reflection, which corresponds to a $10^{-10}$ order of magnitude of the starlight. Therefore we simply ignore the star suppression noise.  Figure \ref{fig:sp} a shows the resultant light curves of the blue band for $\zeta = 90.0^{\circ }$.

We perform the inversion of the light curve for each band. While we take $\zeta$ and $\ThetaS$ as fitting parameters, the estimated values from the single-band photometry is systematically biased a listed in Table \ref{tab:ob} (blue). This is likely due to the anisotropic scattering and variation of clouds. Hence we fix $\zeta$ and $\ThetaS$ to the input values in Figure \ref{fig:sp}. We will discuss about the obliquity determination later. Figure \ref{fig:sp} b-d displays the recovered maps of the blue (panel b), orange (c), and NIR (d) bands.  The recovered maps in all bands exhibit a primary feature of high (low) reflectivity at high (low) latitude. Comparing with the annual mean of the cloud optical depth ( Fig. [\ref{fig:sp}] e), one can interpret the primary feature as the mean cloud distribution. While the spatial resolution of the inversion is too poor to recover the narrow band of persistent clouds seen at the equator, known as the Intertropical Convergence Zone, broad bands of the polar front cloud at high latitude is well recovered in these maps. While short-time scale variations of clouds in weeks or months make systematic residuals between the prediction and data, as indicated in Figure \ref{fig:sp} a, these variations do not affect recovered mean features of clouds so much.

\begin{table}[!tbh]
\begin{center}
\caption{Estimated obliquity $\zeta$ and $\ThetaS$. \label{tab:ob}}
  \begin{tabular}{cc}    
   \hline\hline 
Input & Estimated $^\ast$ \\
\hline
\multicolumn{2}{c}{\bf blue (S/N=20) } \\
$\zeta=90^\circ$, $\ThetaS=0^\circ$ & $\zeta=57.3^\circ \pm 1.0^\circ$, $\ThetaS=12.6^\circ \pm 0.9^\circ$  \\   
\hline
\multicolumn{2}{c}{\bf NIR - blue (S/N=20)} \\
$\zeta=90^\circ$, $\ThetaS=0^\circ$ & $\zeta=90.0^\circ \pm 1.1^\circ$, $\ThetaS=-1.7^\circ \pm 3.2^\circ$  \\   
$\zeta=60^\circ$, $\ThetaS=0^\circ$ & $\zeta=59.7^\circ \pm 3.1^\circ$, $\ThetaS=15.5^\circ \pm 2.9^\circ$  \\
$\zeta=45^\circ$, $\ThetaS=0^\circ$ & $\zeta=50.4^\circ \pm 2.5^\circ$, $\ThetaS=17.1^\circ \pm 3.9^\circ $   \\
$\zeta=30^\circ$, $\ThetaS=0^\circ$ &  $\zeta=38.1^\circ \pm 3.0^\circ$, $\ThetaS=19.8^\circ \pm 3.3^\circ$  \\
$\zeta=23.4^\circ$, $\ThetaS=0^\circ$ & $\zeta=30.1^\circ \pm 2.3^\circ$, $\ThetaS=23.5^\circ \pm 3.3^\circ$ \\
\hline
\multicolumn{2}{c}{\bf NIR - orange (S/N=20)} \\
$\zeta=90^\circ$, $\ThetaS=0^\circ$ & $\zeta=83.5^\circ \pm 3.0^\circ$, $\ThetaS=-2.0^\circ \pm 12.0^\circ$  \\   
\hline
\multicolumn{2}{l}{$\ast \, 0^\circ<\zeta<90^\circ$ and $-180^\circ<\ThetaS<180^\circ$} \\
\multicolumn{2}{l}{The 1 $\sigma$ errors are estimated by the bootstrap resampling. } \\
\end{tabular}
\end{center}
\end{table}

\begin{figure*}[htb]
  \includegraphics[width=\linewidth]{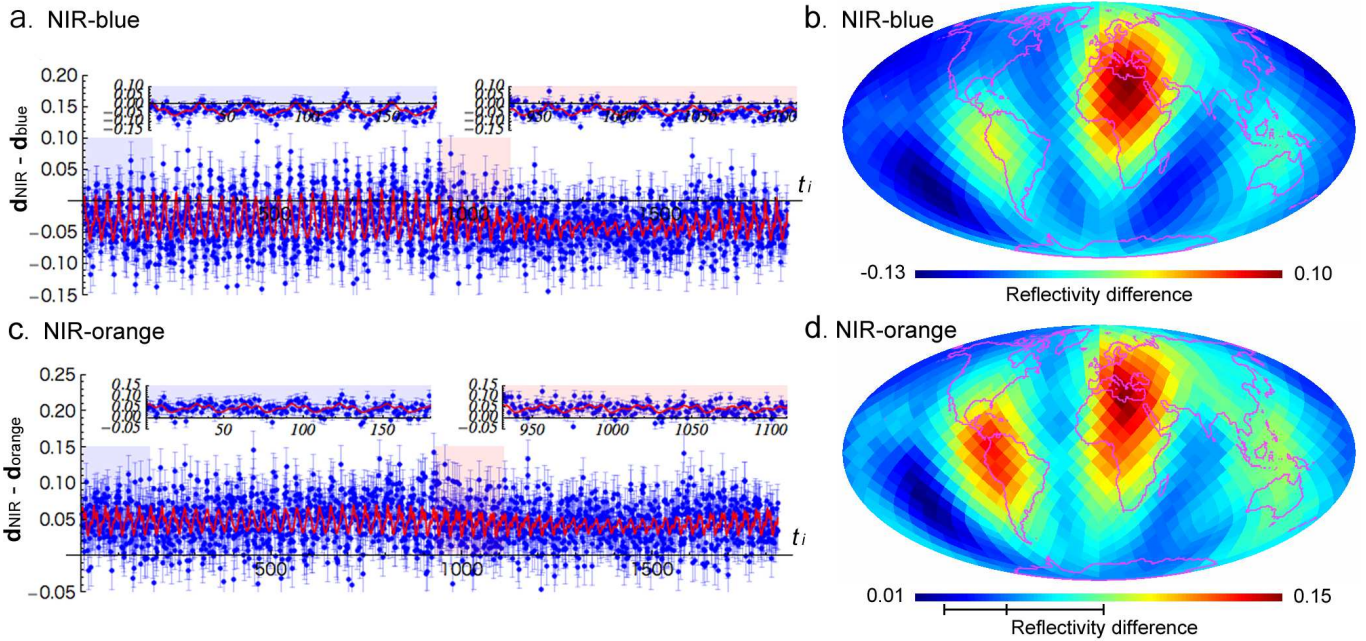}
\caption{The panels a and c display reflectivity differences between the NIR and blue bands and the NIR and orange bands. The recovered maps of reflectivity difference are shown in the panel b (NIR-blue) and d (NIR-orange). In this figure we simultaneously estimate $\zeta$ and $\ThetaS$ listed in Table \ref{tab:ob}. The black bar in the panel d indicates the maximum (0.02), averaged (0.05) , and minimum (0.08) values of the intrinsic reflectivity difference (NIR-orange) of the spatially-unresolved data within a 1 year observation (30 $\times$ 366 bins) .} \label{fig:dif} 
\end{figure*}

Though a primary feature is dominated by clouds, slight differences of the recovered maps between different bands are due to surface components other than clouds. Figure \ref{fig:al} demonstrates several representative examples of the surface reflectivity spectra in 0.4 - 1.0 $\mu$m. The reflectivity of soil and vegetation increases as wavelength increases. This band dependence causes the differences between the recovered maps. Due to almost constant reflectivity of clouds (Fig. [\ref{fig:al}]), the cloud signal is suppressed by taking the difference between two bands. Applying the spin-orbit tomography to the difference vector $\dv_\mathrm{NIR} - \dv_\mathrm{blue}$ (Fig. [\ref{fig:dif}] a), we obtain the NIR-blue map as shown in Figure \ref{fig:dif} b. The contrast standing out in the NIR-blue map traces the approximate continental distribution of the Earth. The NIR-blue map hardly exhibits features at the north America Continent and northern part of the Eurasian Continent due to the large cloud coverage (Fig. [\ref{fig:sp}]) and almost constant reflectivity of snow (Fig. [\ref{fig:al}]). The reason of negative value on the oceans is that Rayleigh scattering dominates the reflectivity.  

The estimated $\zeta$ and $\ThetaS$ agree well with the input value in this case because the variation and the anisotropic effect of clouds are suppressed. We also perform the inversion for different obliquities listed in Table \ref{tab:ob}. The errors of $\zeta$ and $\Theta_S$ are estimated by the bootstrap resampling. We find that the inversion of the NIR-Blue bands can constrain the obliquity, while the estimate of low obliquity tends to be difficult due to less available information and to be slightly biased. 

We also solve the inversion of difference of two bands across the red edge, $\dv_\mathrm{NIR} - \dv_\mathrm{orange}$ (Fig. [\ref{fig:dif}] b). The red edge feature has been examined as a potential biomarker by both Earth-shine observation \citep{woolf2002,arnold2002,2005AsBio...5..372S,rodriguez2005,rodriguez2006,hamdani2006} and simulations \citep{tinetti2006a, tinetti2006b, rodriguez2006, arnold2009}. Compared with the NIR-blue map, the NIR-orange map (Fig. [\ref{fig:dif}] d) displays larger reflectivity near the equator, corresponding to the rain forests in Amazon and Southeast Asia. Inhabited exoplanets are likely to exhibit the localization of photosynthetic organisms because of inhomogeneous insolation and precipitation. The enhancement of the edge-like signature at regions suited for photosynthesis might support the presence of extraterrestrial plants. 

The red edge feature in spatially-unresolved data is generally diminished by contributions of oceans and clouds. The bar in Figure \ref{fig:dif} d indicates the maximum, average, and minimum of the reflectivity difference of spatially-unresolved light curve ($\dv_\mathrm{NIR} - \dv_\mathrm{orange}$) without noise. Since the red edge feature recovered near the tropical regions is 2-3 times larger than the maximum and average of the NIR-orange of the light curve, we conclude that the two-dimensional mapping improves the detectability of the red-edge feature.

In this paper, we have assumed a 4 m telescope, which is smaller than that of other works \citep[e.g.][]{2009ApJ...700..915C}. For instance, \cite{2009ApJ...700..915C} assumed a 16 m telescope with a coronagraph to achieve 2 \% photometry with 1 hour exposures for an Earth twin at 10 pc. Considering the shot noise only, we compute the S/N for 1 hour exposures and a 1 m telescope: 0.32 and 0.44 for their and our fiducial (S/N = 5 \% for 4.8 hour exposures) assumptions. Hence, our noise assumption is similar to them. The main difference in a telescope size requirement comes from that our method uses data in a whole year. We also assumed the spin rotational period, which is a critical parameter for our method. \citet{2008ApJ...676.1319P} suggested that S/N=20 for 0.5 hour exposures in a 8-week observation is enough to measure $\omegaspin$ by the periodogram analysis, corresponding to $\sim$ 3 times larger aperture than ours (i.e. $\sim$ 12 m). While a 1 year observation might improve statistics, it is unclear whether the periodogram works in the same manner when one cannot ignore the orbital revolution. Hence the feasibility of the spin-orbit tomography in cooperation with the spin rotation measurement is still under consideration and will be discussed in a future work.

\section{Summary}
Using simulated scattered light curves from an Earth-twin based on the satellite data, we have demonstrated that surface albedo maps can be recovered from annual and diurnal photometric variations of terrestrial exoplanets partially covered with clouds and in face-on orbits. We have shown that the inversion of the light curve recovers the approximate cloud distribution. After reduction of the cloud signal by taking differences between $0.8-0.9 \mu \mathrm{m}$ and  $0.4-0.5 \mu \mathrm{m}$ (or $0.6-0.7 \mu \mathrm{m}$) , we have found that the recovered maps trace continental distribution of the Earth. Moreover, the obliquity and equinox can be estimated simultaneously with the reflectivity difference mapping. 

Though we only consider a face-on orbit in this paper, it is essential to extend our technique to general geometry for practical use. Especially for low-obliquity planets like the Earth, larger area can be recovered in some inclined geometry. For the extension to general inclination, the effect of anisotropic scattering to the phase curve cannot be ignored. We will include these effects to the spin-orbit tomography and apply it to various geometrical cases including transiting planets, low obliquity planets with proper inclination, and even tidally locked planets in a forthcoming paper (Fujii and Kawahara in preparation).

We are deeply grateful to Yasushi Suto, Atsushi Taruya, Eiichiro Kokubo and Edwin L. Turner for helpful and insightful discussion. We are greatly thankful to C. Pichon, who kindly instructed us the basic knowledge of the inverse problem. We also thank an anonymous referee for a lot of constructive comments. The authors are grateful to OpenCLASTR project for using the Rstar6b package in this research. HK is supported by a JSPS (Japan Society for Promotion of Science) Grant-in-Aid for science fellows. This work is also supported by Grant-in-Aid for Scientific research from JSPS and from the Japanese Ministry of Education, Culture, Sports, Science and Technology (Nos. 22$\cdot$5467 and 23$\cdot$6070).

\end{document}